\documentclass[twocolumn,showpacs,floatfix,aps,prl]{revtex4}

\usepackage{graphicx}% Include figure files
\usepackage{dcolumn}% Align table columns on decimal point
\usepackage{bm}% bold math

\begin{document}

\title{Observation of stochastic resonance in percolative Josephson media}

\author{A. M. Glukhov}
\email{glukhov@ilt.kharkov.ua}
\affiliation{B. Verkin Institute
for Low Temperature Physics and Engineering, 61103 Kharkov,
Ukraine}
\author{A. G. Sivakov}
\affiliation{B. Verkin Institute for Low Temperature Physics and
Engineering, 61103 Kharkov, Ukraine}
\author{A. V. Ustinov}
\affiliation{Physikalisches Institut III, Universit\"at
Erlangen-N\"urnberg, D-91058 Erlangen, Germany}

\date{\today}

\begin{abstract}
Measurements of the electrical response of granular Sn-Ge thin
films below the superconducting transition temperature are
reported. The addition of an external noise to the magnetic field
applied to the sample is found to increase the sample voltage
response to a small externally applied ac signal. The gain
coefficient for this signal and the signal-to-noise ratio display
clear maxima at particular noise levels. We interpret these
observations as a stochastic resonance in the percolative
Josephson media which occurs close to the percolation threshold.
\end{abstract}

 \maketitle

\subsection{1. Introduction}

The phenomenon of \emph{stochastic resonance} has been discussed
in relation to diverse problems in nonlinear science, physics,
chemistry and biology \cite{Hanggi}. Generally speaking,
stochastic resonance is the enhancement of the output
signal-to-noise ratio caused by the injection of an optimal
amount of noise into a periodically driven nonlinear system. This
kind of behavior is often thought as counterintuitive, since here
a stochastic force amplifies a small periodic signal. Its
mechanism is usually explained in terms of motion of a particle
in a double-well potential subjected to noise, in the presence of
a time-periodic force. The periodic forcing leads to
noise-enhanced transitions between the two wells and thus to an
enhanced output of the forcing signal.

A clear example of nonlinear systems with few degrees of freedom
is a superconducting loop with a Josephson junction, well known
as a superconducting quantum interferometer (SQUID). With a
proper choice of the size of the loop, this system undergoes
bistable dynamics for magnetic flux trapped in the loop. There
have already been experiments that reported operating SQUIDs under
stochastic resonance conditions, both with external noise
injection \cite{Hibbs} and with thermally generated intrinsic
noise \cite{Lukens}. The stochastic resonance effect can be
considerably enhanced in a system of coupled bistable oscillators
(see, e.g., Ref.~\cite{Bulsara}). Therefore, it is interesting to
study stochastic amplification for a Josephson media consisting
of many superconducting contours with Josephson junctions.

Earlier we observed quantum interference effects in
macroscopically inhomogeneous superconducting Sn-Ge thin-film
composites near the percolation threshold \cite{PhysicaB}. This
system exhibits a considerable voltage noise under dc current
bias and a rectification of ac current, which arise below the
superconducting transition temperature. According to
Ref.~\cite{Ouboter}, a dc voltage is observed when an ac current
larger than the critical current passes through a system of two
superconductors weakly connected by an asymmetric double point
contact, i.e., the magnetic flux quantization induces the critical
current oscillations and the respective voltage oscillations. We
have argued \cite{PhysicaB}, that the oscillatory dependence
$V_{dc}(H)$ in Sn-Ge thin-film composites is related to quantum
interference in randomly distributed asymmetric superconducting
contours interrupted by Josephson \emph{weak links}.

In Ref.~\cite{PhysicaB} we reported measurements of the
$V_{dc}(H)$ dependence for various orientations of the film
relative to the field. The scale of the oscillatory structure in
$V_{dc}(H)$ is inversely proportional to the cosine of the angle
between the applied magnetic field and the normal to the sample
plane. The emergence of the normal magnetic field component alone
and also the antisymmetry of the oscillatory structure relative to
$H=0$ indicate a quantum-interference origin of $V_{dc}(H)$.
Moreover, it appears feasible to relate these active contours to
the percolative cluster that has a well-known \emph{fractal}
structure. The existence of a wide and \emph{self-similar}
distribution of Josephson contour areas leads to a fractal
character of the dependence $V_{dc}(H)$. We have suggested and
verified a model for the origin of the $1/f$ voltage noise by a
passive transformation of magnetic field oscillations with a
fractal transfer function $V_{dc}(H)$~\cite{PhysicaB}.

In the present paper, we study the noise-induced electrical
response of granular Sn-Ge thin-film composites. We argue that a
distributed network containing many superconducting loops with
Josephson junctions may show a cooperative behavior as
\emph{stochastically resonating media}.

\subsection{2. Experimental details and results}

\begin{figure}
\includegraphics{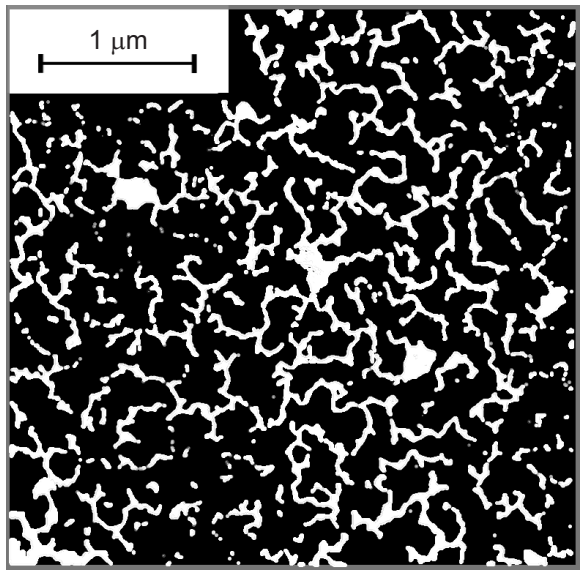}
\caption{Electron micrograph of Sn-Ge sample prepared close to the
percolation threshold. Black regions correspond to the metallic
phase.} \label{micrograph}
\end{figure}

Josephson networks may occur naturally, e.g., in nonuniform
superconducting materials such as granular thin films. We prepare
granular Sn-Ge thin-film composites having monotonically varying
structure by vacuum condensation of Sn on a long (60 mm)
substrate along which a temperature gradient is created. Sn is
deposited on the previously prepared 50 nm thick Ge layer. The
thickness of the Sn layer is 60 nm. The metallic condensate is
covered from the top with amorphous Ge. The structural change
results in variation of the composite properties from metallic to
insulating over the substrate. This crossover in properties is
observed on a series consisting of 30 samples cut from different
parts of the substrate. For the present investigations, we chose
samples with properties near the percolation threshold, with a
characteristic structure depicted in Fig.~\ref{micrograph}.

During the measurements, the samples were kept in exchange gas
inside a superconducting solenoid. The electrical measurements
were carried out according to the standard four-probe technique. A
sinusoidal ac current of frequency $f_{I}=100$ kHz and amplitude
$I_{ac}=0.8$ mA was produced by an HP3245A universal source
connected to the current leads through a dc-decoupling
transformer. Fast Fourier transformation spectra of the output
voltage are measured by using an SR770 spectrum analyzer with a
Blackman-Harris window function. We used the signal-to-noise
ratio (SNR) as the major characteristic of stochastic resonance.
The SNR was measured as the ratio of the voltage amplitude of the
spectral line to the voltage noise level below it. The noise
background in the signal bin is estimated by performing a linear
fit to the peak clipped spectrum. The noise intensity (noise
level) denotes the standard deviation $\sigma_{N}$ of the Gaussian
white noise signal, which was supplied by the internal SR770
generator.

\begin{figure}
\includegraphics{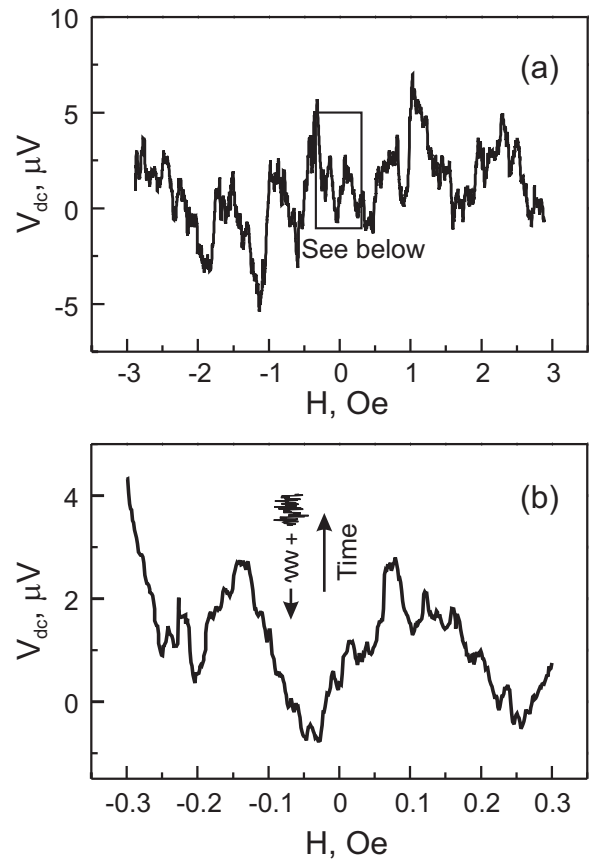}
\caption{(a) Oscillatory behavior of the rectified voltage across
the Sn-Ge sample versus dc magnetic field: $T=3.0$ K, $f_{I}=100$
kHz and $I_{ac}=0.8$ mA. (b) Illustration of the stochastic
resonance detection scheme. Magnetic field components $H_{ac}$ and
$H_{noise}$ are added to dc magnetic field $H$.} \label{VdcH}
\end{figure}

The transition of a sample into the superconducting state is
smeared over 1.0 K with the center of the resistive transition at
$T_{0}=3.8$ K. At temperatures below $T_{0}$ and with ac current
$I_{ac}$ applied through the sample, we observed a rectified dc
voltage $V_{dc}$, the magnitude of which oscillated as a function
of the dc magnetic field $H$ applied perpendicular to the
substrate (Fig.~\ref{VdcH}a). The amplitude and frequency of the
current $I_{ac}$ did not significantly affect the general
features of the $V_{dc}(H)$ dependence. The results could be
always readily reproduced.

\begin{figure}
\includegraphics{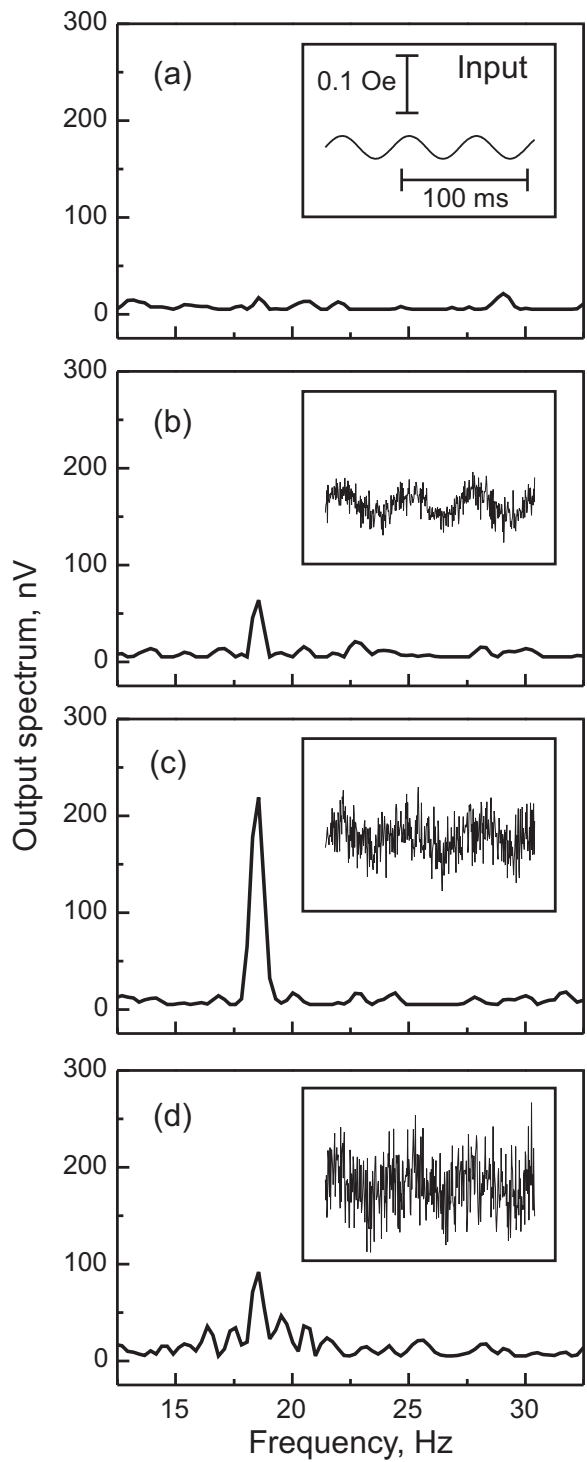}
\caption{Input signal $H_{ac}+H_{noise}$ (insets) and the Fourier
spectrum of the output voltage for different levels of input
noise $\sigma_{N}$, mOe: 0 (a), 16 (b), 31 (c), 47 (d). The input
signal amplitude remains constant at $H_{ac}=20$ mOe. Signal
frequency $f_{H}=18.5$ Hz, dc magnetic field $H=0.17$ Oe.}
\label{FFT}
\end{figure}

\begin{figure}
\includegraphics{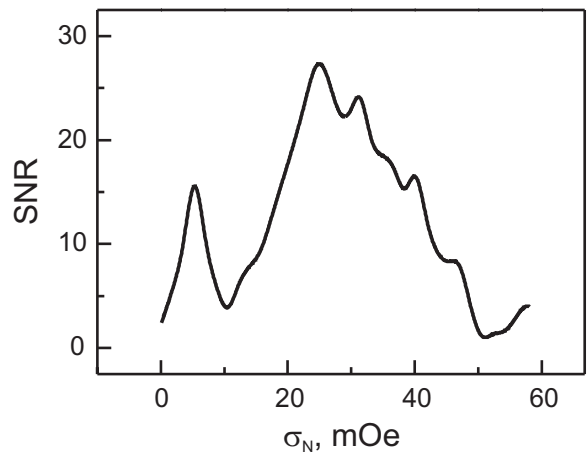}
\caption{Output signal-to-noise ratio (SNR) versus input noise
level $\sigma_{N}$ for the first harmonic of the input signal
frequency $f_{H}=18.5$ Hz. Magnetic field $H=0.17$ Oe.}
\label{SNR}
\end{figure}

\begin{figure}
\includegraphics{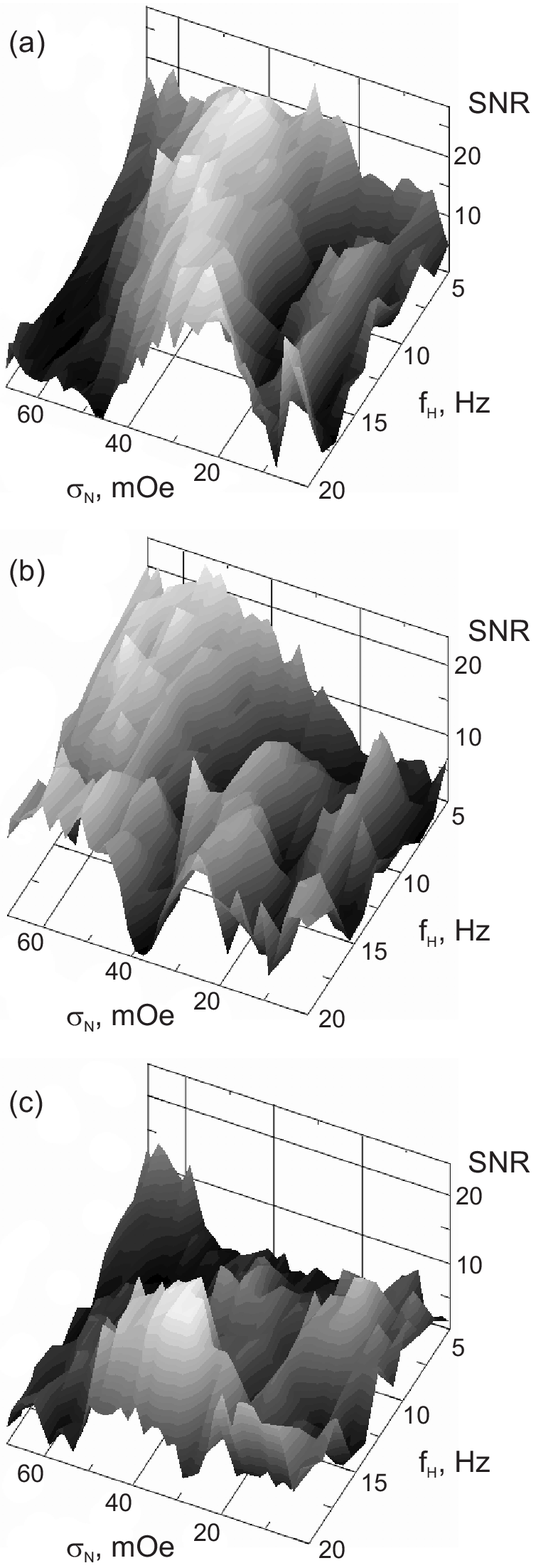}
\caption{SNR dependence on input noise level $\sigma_{N}$ and
input signal frequency $f_{H}$ at different dc magnetic fields
$H$, Oe: 0.17 (a), 0.18 (b), 0.19 (c).} \label{3D_SNR}
\end{figure}

To observe the phenomenon of stochastic resonance, we study the
rectified voltage dependence on magnetic field. The applied
magnetic field consisted of three components (Fig.~\ref{VdcH}b):
(i) a dc field $H$, which varied in the range between $-300$ and
$+300$ mOe, (ii) a small ac component with a frequency $f_{H}$
between 5 and 60 Hz and an amplitude $H_{ac}=20$ mOe, and (iii)
Gaussian white noise $H_{noise}$ with the intensity $\sigma_{N}$
ranging up to 70 mOe. The Fourier spectra of voltage response are
shown in Fig.~\ref{FFT} together with oscillograms of the input
signal $H_{ac}+H_{noise}$. Figure~\ref{SNR} shows the dependence
of the output SNR for the first harmonic of $f_{H}$ on the
intensity of input noise $H_{noise}$. One can see that increasing
the noise amplitude at first increases the SNR and then decreases
it. Such maxima are rather characteristic for the phenomenon of
stochastic resonance. Similar measurements taken at different
magnetic fields and frequencies often showed multiple maxima such
as those shown in Fig.~\ref{3D_SNR}.

\subsection{3. Discussion}

In summary, our experiments demonstrate the characteristic
feature of the phenomenon of stochastic resonance, namely the
nonmonotonic behavior of the SNR. At the optimum noise level the
SNR increases up to 40. The presence of multiple maxima
(Fig.~\ref{SNR} and~\ref{3D_SNR}) can be due to the effect of
different Josephson contours in our structure, which is operated
at the border of the percolation threshold.

We suppose that the nonmonotonic dependence of the SNR on
frequency $f_{H}$ (Fig.~\ref{3D_SNR}) excludes other possible
explanations (such as, e.g., a simple rectification effect due to
a nonlinearity of the response) for the observed gain of a small
input signal.

Detailed measurements taken at different frequencies, shown in
Fig.~\ref{3D_SNR} indicate, at least in some ranges of the dc
magnetic field, the existence of parameter regions characterized
by a significant gain for a relatively broadband signal. We
interpret this behavior as a property of percolative Josephson
media with a wide range of self-similar loops. The SNR gain in our
system can be tuned to a desired operation frequency $f_{H}$ by
changing the dc magnetic field H.

The nature of the stochastic resonance in the system studied can
be related to the commonly known bistable oscillator behavior of
the magnetic flux quantization loops. Moreover, in the presence of
current bias $I_{ac}$ at relatively high frequency (at $f_{I}$
about 100 kHz) with amplitude larger than critical, our samples
exhibit dynamical chaos. Such a regime is commonly characterized
by a coexistence of multiple attractors in the phase space.
Indeed, calculation of Lyapunov exponents from the time evolution
of the voltage measured at constant current indicates presence of
chaos in our system~\cite{LT22}. In this case, the "phase
trajectory" of the system may stay long time in any of the
attractors and perform irregular transitions between them.
Synchronization of such intermittent transitions by a small input
signal may lead to stochastic resonance as
well~\cite{Anischenko}. Yet, these speculations require further
investigations to be firmly justified.

This work was supported by the German-Ukrainian collaboration
grant of the Bundesministerium f\"{u}r Bildung, Wissenschaft,
Forschung und Technologie (BMBF) No. UKR-003-99.

\end{document}